\documentclass[a4paper,11pt]{article}
\usepackage{pos}

\title{New results on inclusive $B\to X_{u} \ell \nu$ decay from the Belle experiment}

\author*[1]{Lu Cao}

\affiliation{University of Bonn,\\
  Nussallee 12, Bonn, Germany}

\note{On behalf of the Belle collaboration.}

\emailAdd{cao@physik.uni-bonn.de}

\abstract{We report on the measurement of inclusive charmless semileptonic B decays $B \to X_{u} \ell \nu$. The analysis makes use of hadronic tagging and is performed on the full data set of the Belle experiment comprising 772 million $B\bar{B}$ pairs. In the proceedings, the preliminary results of measurements of partial branching fractions and the CKM matrix element $|V_{ub}|$ are presented.}

\FullConference{%
  40th International Conference on High Energy physics - ICHEP2020\\
  July 28 - August 6, 2020\\
  Prague, Czech Republic (virtual meeting)
}

\begin{document}
\maketitle

\section{Introduction}
In the Standard Model of particle physics (SM), the Cabibbo-Kobayashi-Maskawa (CKM) matrix \cite{Cabibbo:1963yz, Kobayashi:1973fv} describes the quark mixing and accounts for $CP-$violation in the quark sector. One of the crucial tests of the SM is precise determination of the magnitude of the matrix elements. In $b-$flavor scope, the corresponding world averages of $|V_{ub}|$ from both exclusive and inclusive determinations \cite{Amhis:2019ckw} are 
\begin{equation}
\label{eq:avr-vub}
\begin{array}{l}
\left|V_{u b}^{\text {excl. }}\right|=(3.67 \pm 0.09 \pm 0.12) \times 10^{-3} ,\\
\left|V_{u b}^{\text {incl. }}\right|=\left(4.32 \pm 0.12_{-0.13}^{+0.12}\right) \times 10^{-3} ,
\end{array}
\end{equation}
where the uncertainties are from experiment and theory. The disagreement between them is about three standard deviations.

\begin{figure}[h]
	\centering
	    \includegraphics[width=0.49\linewidth]{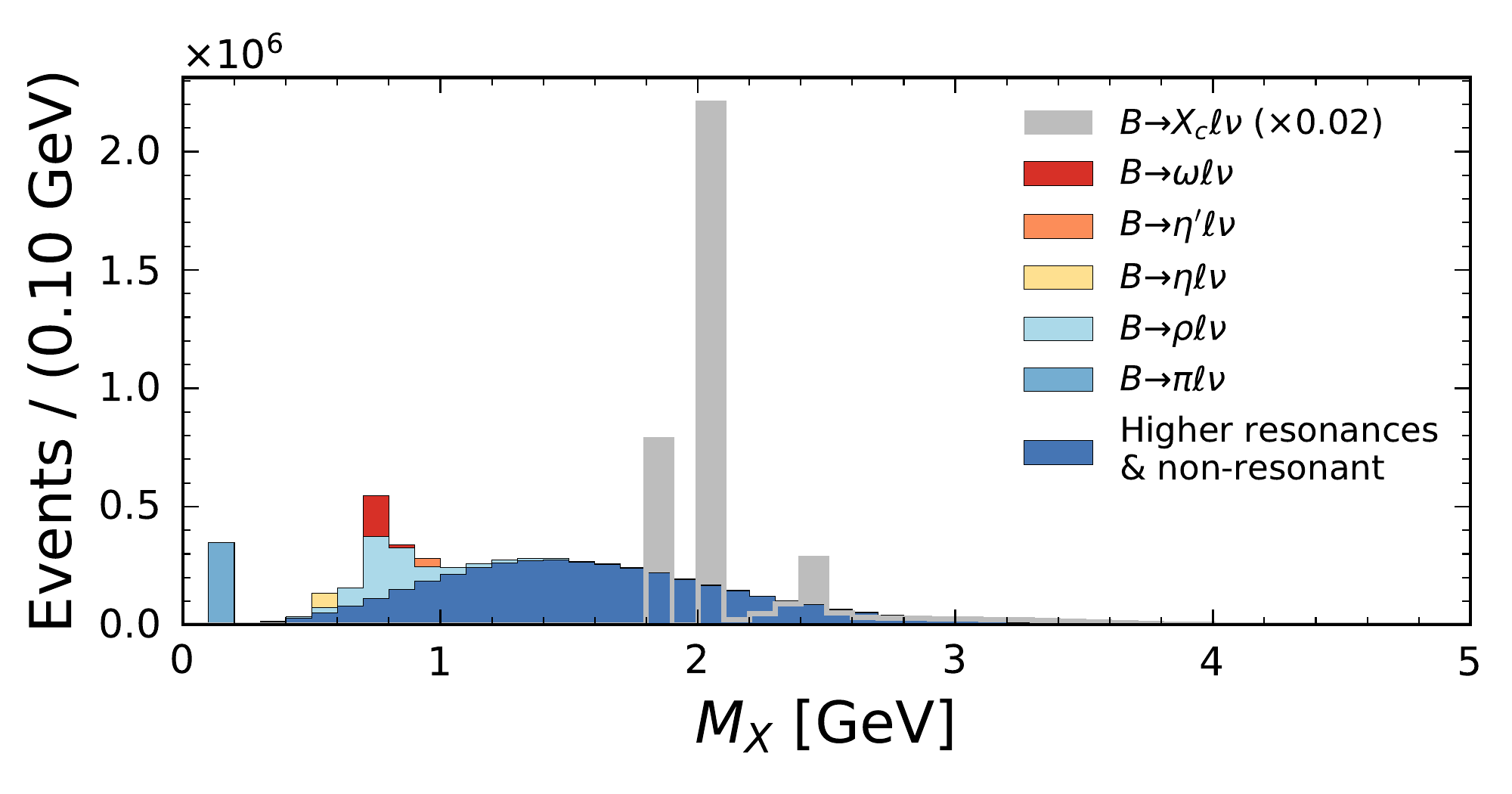}
	    \includegraphics[width=0.49\linewidth]{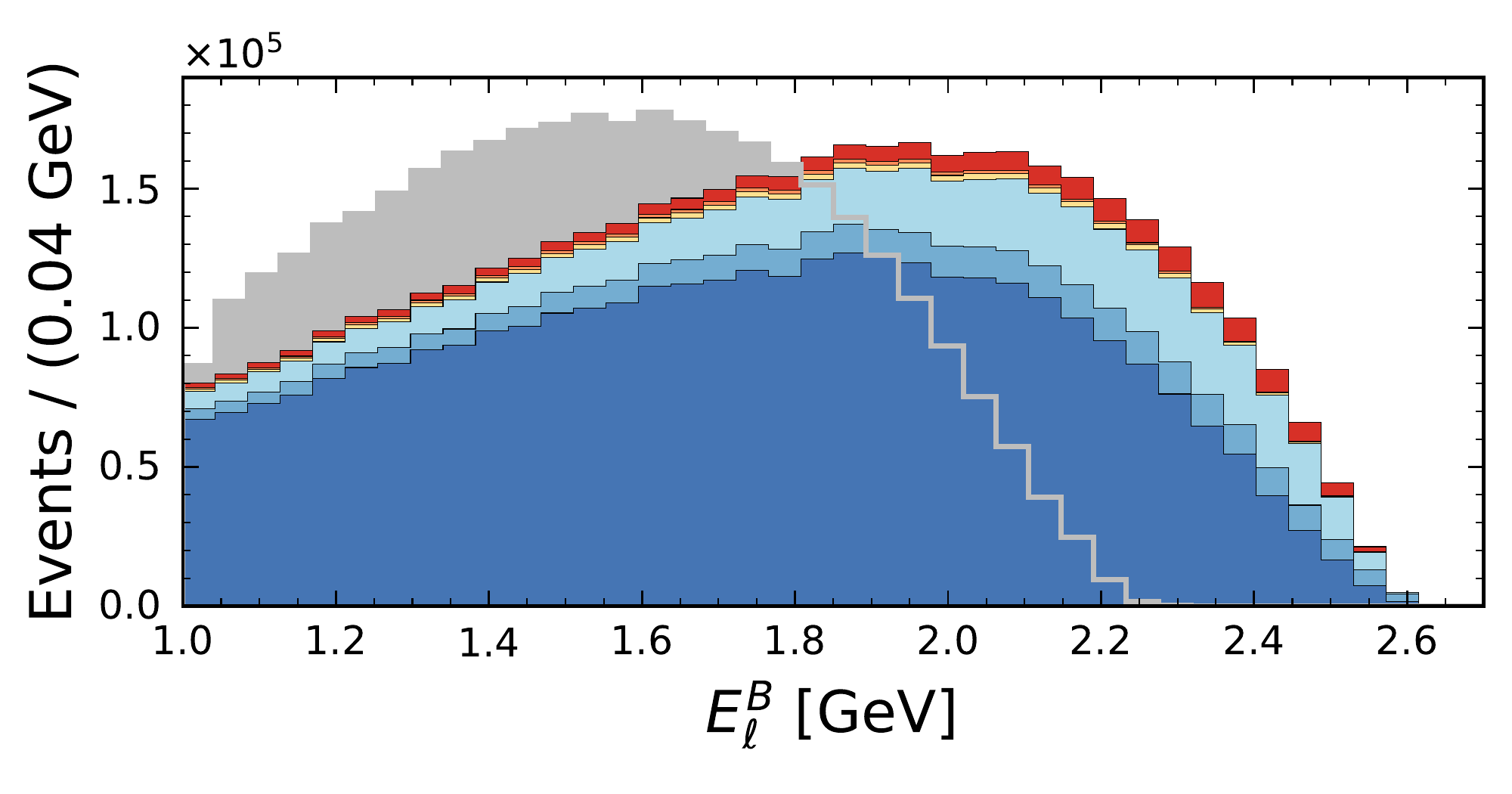}
	\caption{\label{fig:gen} The generator-level $M_{X}$ and $E_{\ell}^{B}$ distributions of the $B\to X_{u} \ell \nu$ decay comparing to that of $B\to X_{c} \ell \nu$ decay. The $B\to X_{c} \ell \nu$ component (gray) is scaled down by a factor of 50 for illustration.}
\end{figure}

On the other hand, the experimental measurement of the inclusive semileptonic decay $B\to X_{u} \ell \nu$ is challenging due to the large background from the CKM-favoured $B\to X_{c} \ell \nu$ decay. Fig.~\ref{fig:gen} illustrates the $B\to X_{u} \ell \nu$ and $B\to X_{c} \ell \nu$ decays with the generator-level distributions in two important kinematic variables: the invariant mass of hadronic system $M_{X}$ and the lepton energy in the signal $B$ rest frame $E_{\ell}^{B}$. It's shown that the clear separation of the signal decay is only possible in certain kinematic regions, e.g. the endpoint of lepton energy or the low $M_{X}$ region. The details of the reconstruction and separation strategy is described in Sec.~\ref{sec:analysis}. The preliminary results on the measured partial branching fractions and the $|V_{ub}|$ values are presented in Sec.~\ref{sec:results}.

\section{Analysis strategy}
\label{sec:analysis}
The data used in this analysis were recorded with the Belle detector \cite{Abashian:2000cg} at the KEKB accelerator complex \cite{KEKB} with a center-of-mass energy of $\sqrt{s} = 10.58$ GeV. The full data set contains an integrated luminosity of 711 fb$^{-1}$ and corresponds to 772 million $\Upsilon(4S) \to B\bar{B}$ events. The Monte Carlo (MC) simulated events are generated by EVTGEN \cite{EvtGen} and the detector response is modeled using GEANT3 \cite{Geant3}. The signal $B\to X_{u} \ell \nu$ MC sample is a combination of resonances and non-resonant decay using a hybrid modelling approach \cite{hybrid, Prim:2019gtj}. The non-resonant component is based on the theory calculation of Ref.~\cite{DeFazio:1999ptt} with the model parameters in the Kagan-Neubert scheme from Ref.~\cite{Buchmuller:2005zv}. 

The hadronic decays of one of the $B$ mesons are reconstructed via the full reconstruction algorithm \cite{Feindt:2011mr} based on neural networks. In total, over 1104 decay cascades are considered and reconstructed. The efficiencies for charged and neutral $B$ mesons are $0.28\%$ and $0.18\%$, respectively \cite{Bevan:2014iga}. The output classifier score of this algorithm presents the quality of the reconstructed candidates. We select the best candidate of $B_{\texttt{tag}}$ for each event. In addition, we require the beam-constrained mass $M_{\texttt{bc}} = \sqrt{(\sqrt{s}/2)^{2} - |\mathbf{p}_{\texttt{tag}}|^{2}} > 5.27$ GeV to suppress continuum processes ($e^{+}e^{-}\to q\bar{q}$, $q=u,d,s,c$) and beam background.

All tracks and clusters not used in the construction of the $B_{\texttt{tag}}$ candidate are used to reconstruct the signal side. With the fully reconstructed four-momentum of $B_{\texttt{tag}}$ and the known beam-momentum, the signal $B$ rest frame can be defined as 
\begin{equation}\label{eq:psig}
p_{\mathrm{sig}}=p_{e^{+}e^{-}}-\left(\sqrt{m_{B}^{2}+\left|\mathbf{p}_{\mathrm{tag}}\right|^{2}}, \mathbf{p}_{\mathrm{tag}}\right).
\end{equation}

\noindent The signal lepton with $E_{\ell}^{B}=\left|\mathbf{p}_{\ell}^{\mathbf{B}}\right|>1$ GeV is used to identify the semileptonic decays. Here the small correction of the lepton mass term to the energy of the lepton is neglected. A veto on lepton-pair mass is applied to reject the lepton from $J/\psi$ decay and photon conversions. In addition, the charge of lepton is required to be opposite to $B_{\texttt{tag}}$ for the charged $B$ case. With the signal lepton selected, the four-momentum of hadronic system $p_{X}$ is defined as a sum of the four-momenta of tracks and clusters which are not involved in reconstructing the $B_{\texttt{tag}}$ and signal lepton. Furthermore, we reconstruct the missing mass squared and the four-momentum transfer squared $q^{2}$ as

\begin{equation}\label{eq:mm2-q2}
\mathit{MM}^{2}=\left(p_{\mathrm{sig}}-p_{X}-p_{\ell}\right)^{2}, \;\; q^{2}=\left(p_{\mathrm{sig}}-p_{X}\right)^{2}.
\end{equation}

We utilise a machine learning based classification with boosted decision trees (BDTs) to separate the signal $B\to X_{u} \ell \nu$ decay from the background events which are dominated by $B\to X_{c} \ell \nu$. The feature variables used for training include $\mathit{MM}^{2}$, the number of charged kaons and $K_{s}^{0}$, the total charge of event, the vertex fit $\chi^{2}/\texttt{dof}$ between the hadronic system and signal lepton, and the $\mathit{MM}^{2}$ and angular information of a partially reconstructed $B \to D^{*} \ell \nu, D^{*} \to D \pi_{\texttt{slow}}$ decay with the slow pions candidates, where $p^{\texttt{cms}}_{\pi_{\texttt{slow}}}<0.22$ GeV. Due to the small difference between the masses of $D$ and $D^{*}$, the flight directions of the $\pi_{\texttt{slow}}$ and $D^{*}$ are strongly correlated and we estimate the energy of $D^{*}$ as $E_{D^{*}} \approx m_{D^{*}} \times E_{\pi_{\texttt{slow}}}/ (m_{D^{*}} - m_{D})$. On the BDT classifier output, we choose a selection criteria that reject $98.1\%$ of $B\to X_{c} \ell \nu$ decays and retain $24.8\%$ of $B\to X_{u} \ell \nu$ signal decays. The selection efficiency on data is $2.3\%$. 

In addition, the $B_{\texttt{tag}}$ reconstruction efficiency is calibrated using a data-driven approach described in Ref.~\cite{Glattauer:2015teq}. The uncertainty of calibration is considered in systematics. We also apply a continuum efficiency correction to the simulated sample by comparing the difference to the number of reconstructed off-resonance events in data.

\section{Partial branching fractions and $|V_{ub}|$ results}
\label{sec:results}
\begin{figure}[t]
	\centering
	    \includegraphics[width=0.32\linewidth]{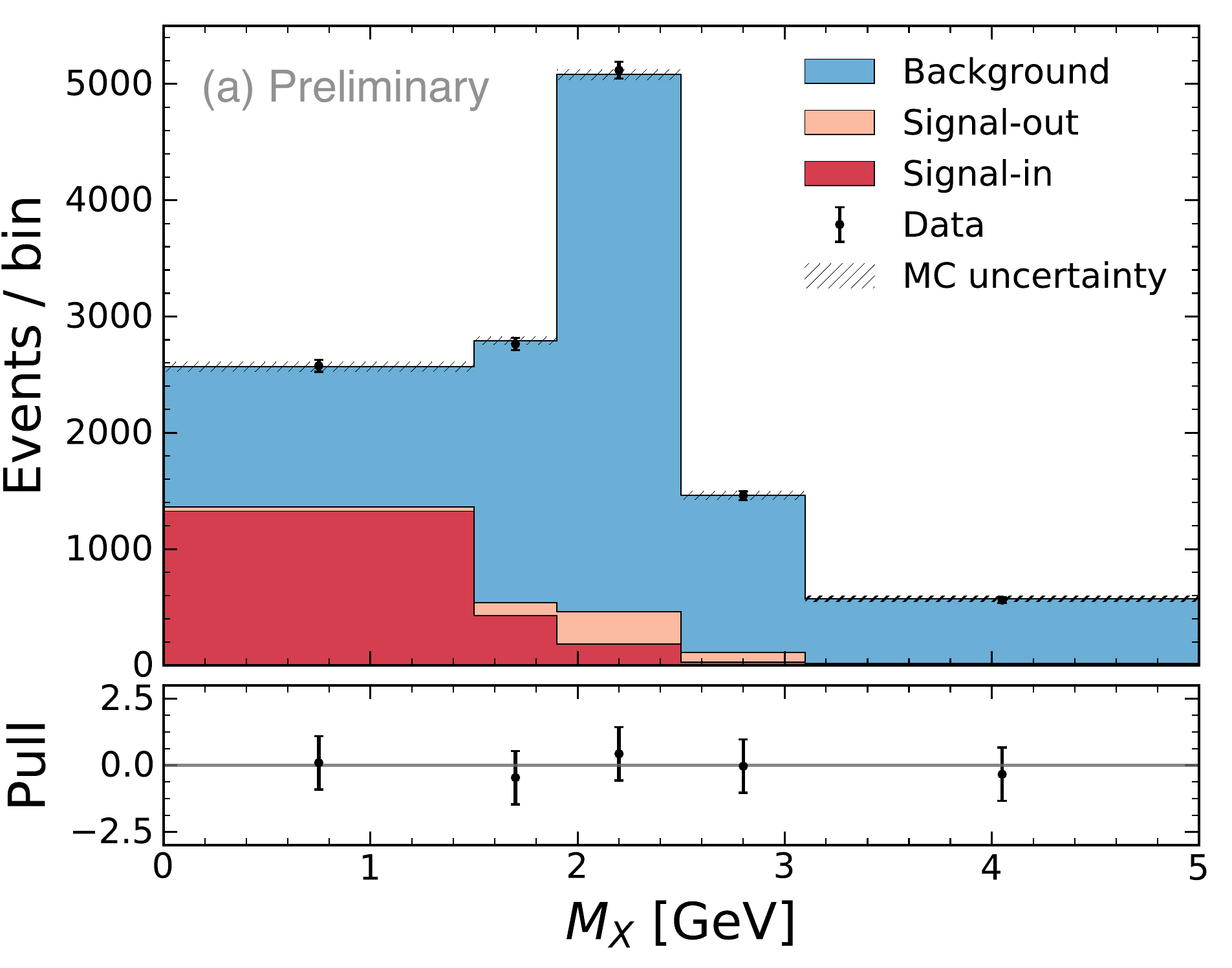}
	    \includegraphics[width=0.32\linewidth]{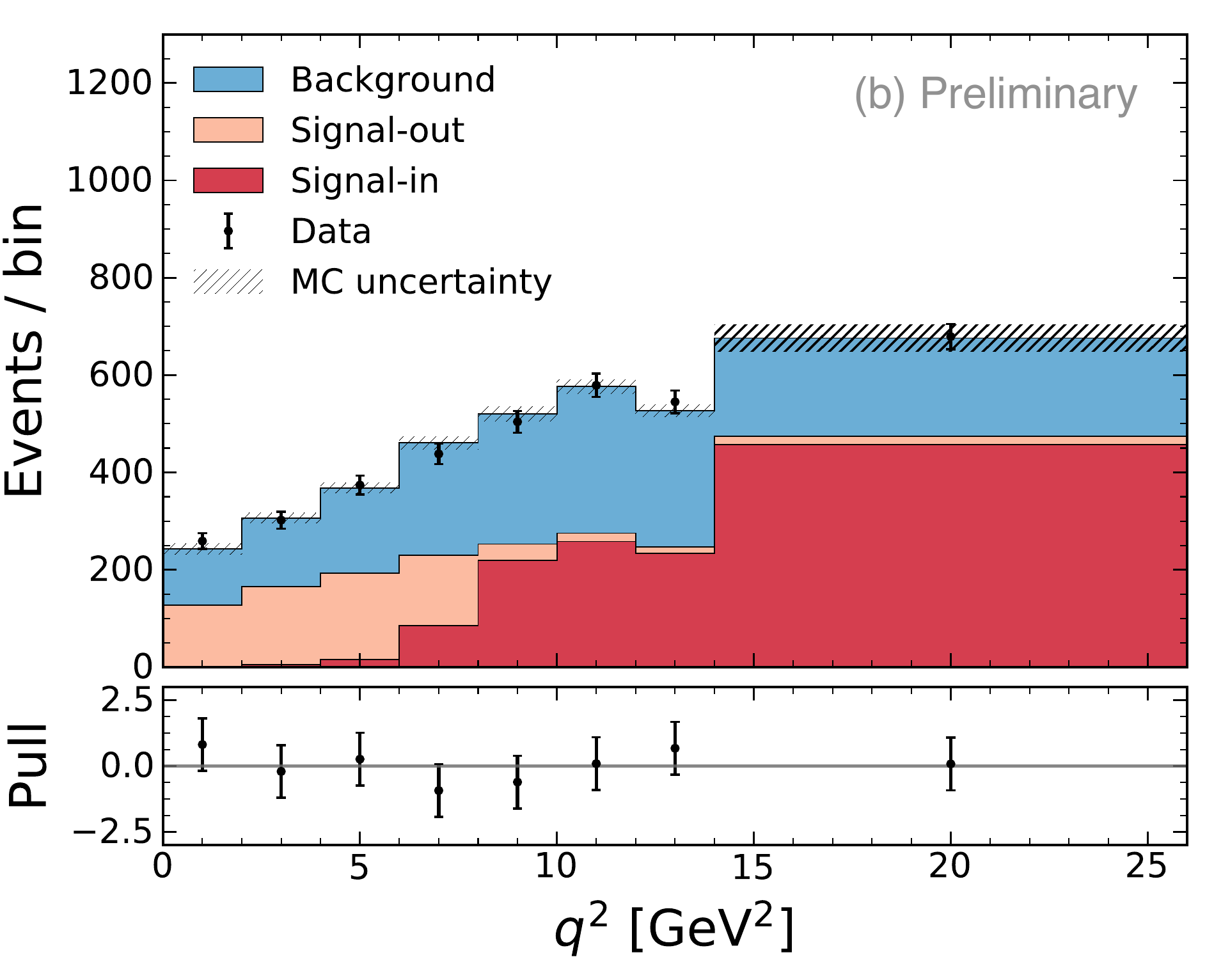}
	    \includegraphics[width=0.32\linewidth]{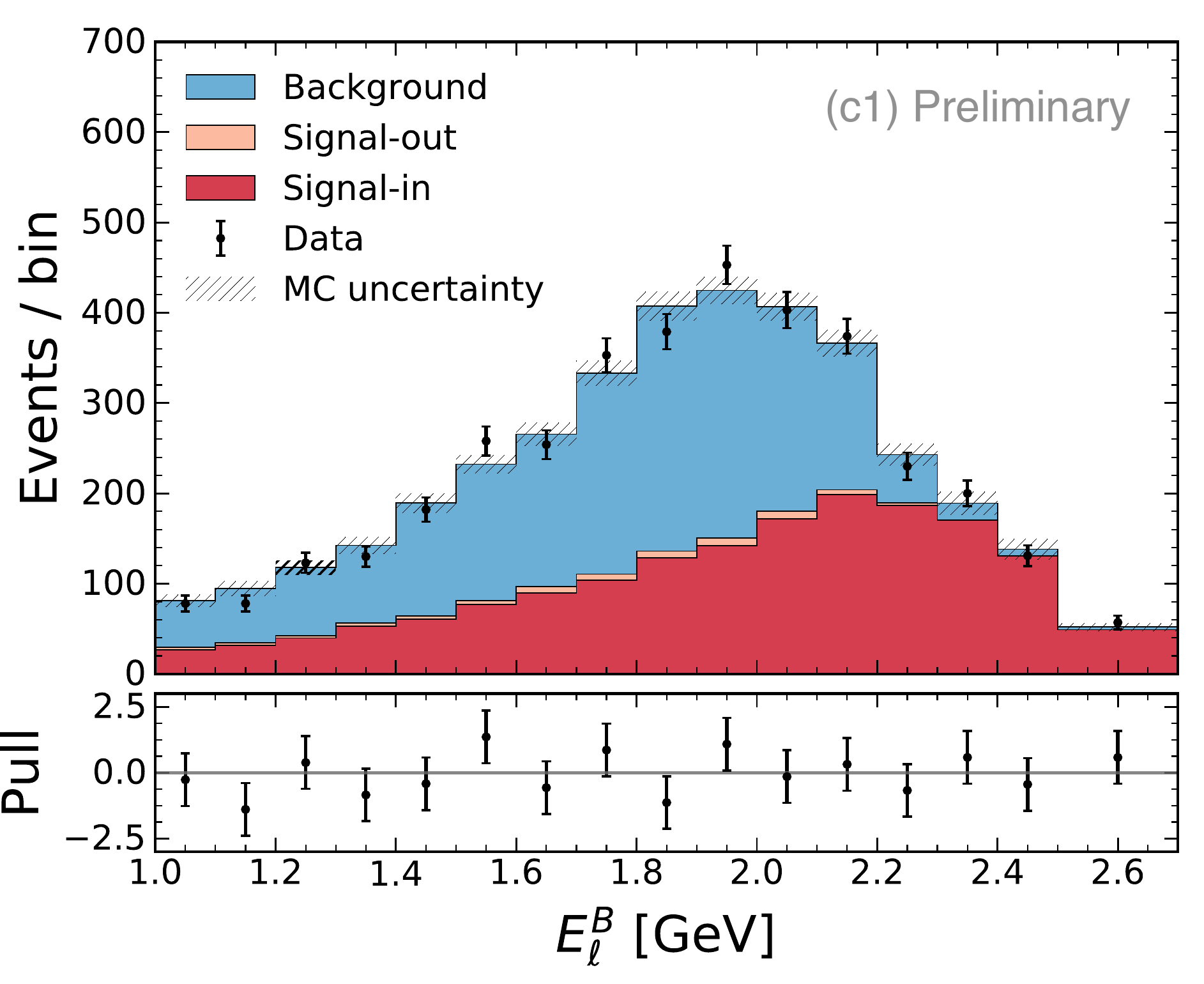}
	    \includegraphics[width=0.32\linewidth]{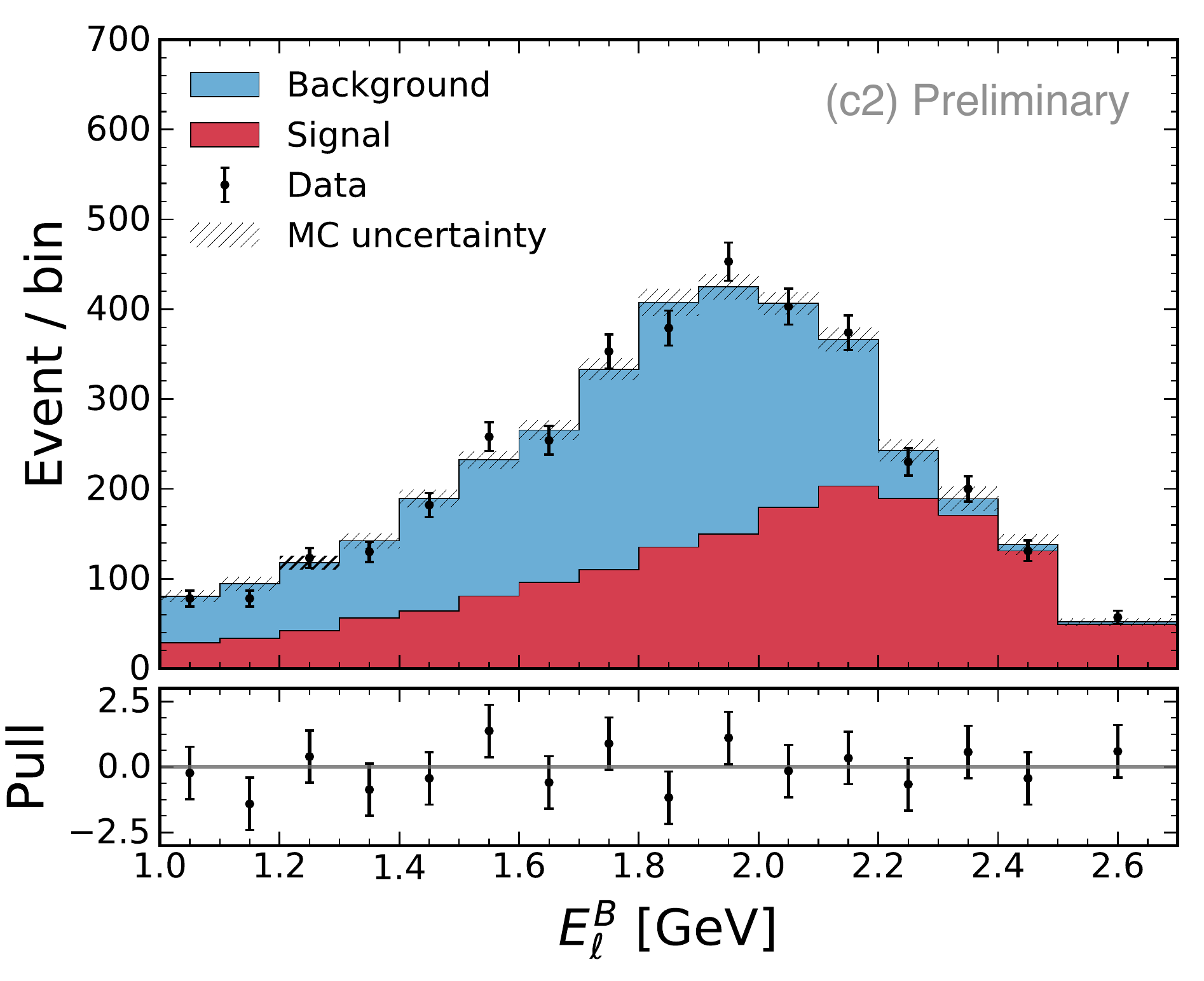}
	    \includegraphics[width=0.32\linewidth]{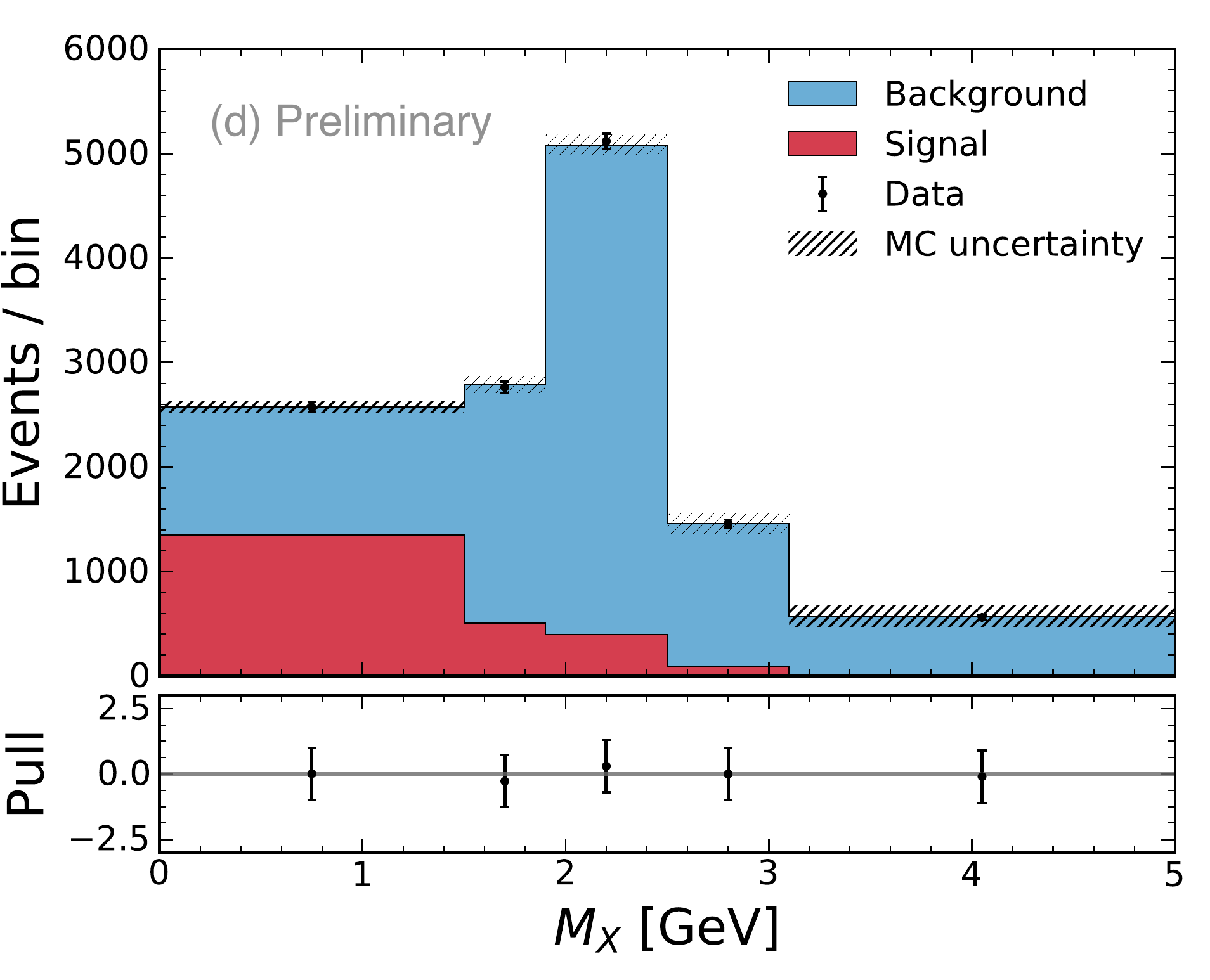}
	    \includegraphics[width=0.32\linewidth]{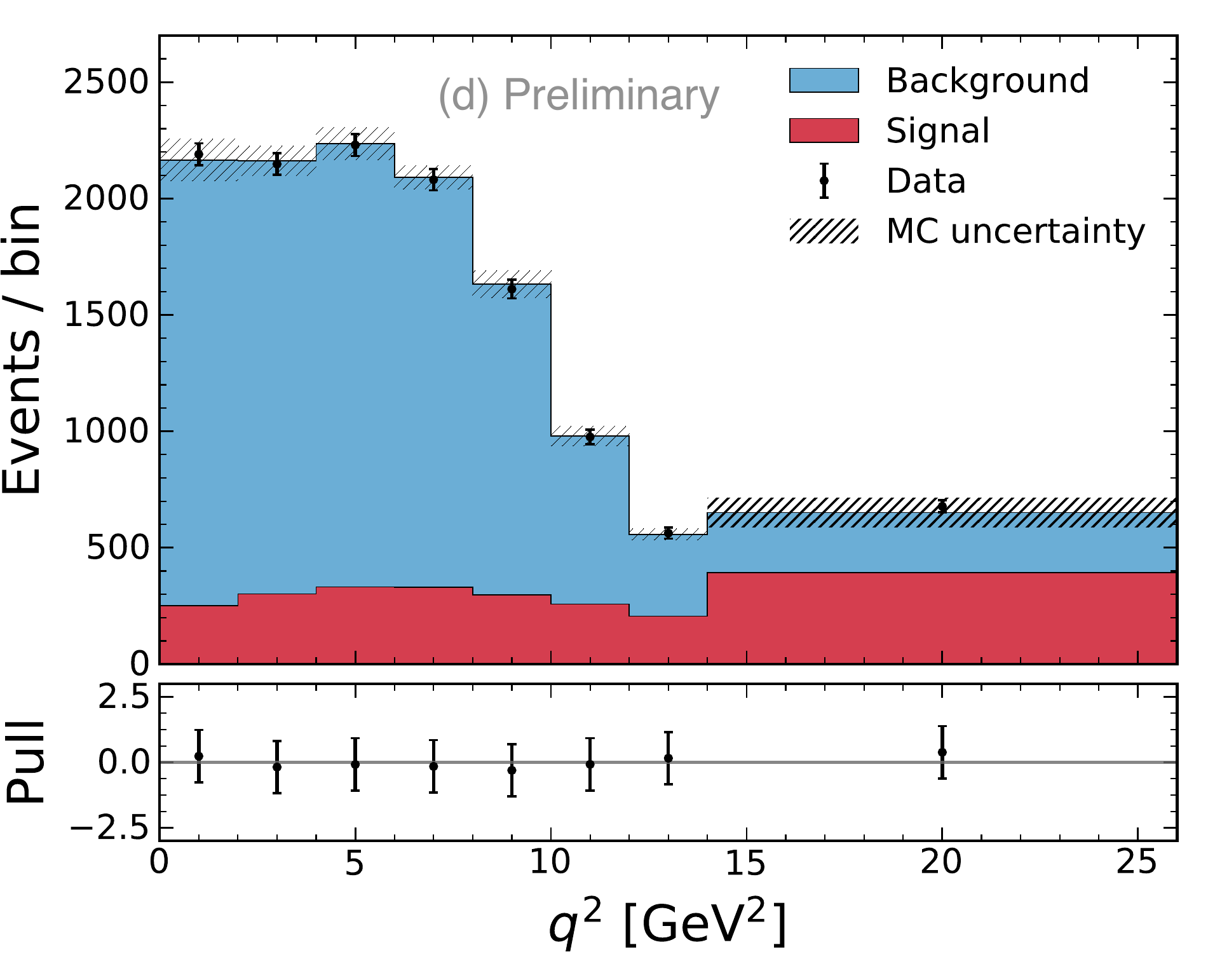}
	\caption{\label{fig:post-fit} The post-fit distributions for various phase-space regions and kinematic variables. The distributions of the two-dimensional fit (d) are shown on the projections of $M_{X}$ and $q2$.}
\end{figure}

\begin{table}[b]
\renewcommand\arraystretch{1.0}
\centering
\begin{tabular}{lcccc}
\hline\hline

Fit & Fit variable       & Phase-space region                                                                           & $10^{3}\Delta \texttt{BF}$  \\ \hline
(a)    &$M_{X}$       & $E^{B}_{\ell}>1$ GeV, $M_{X}<1.7$ GeV                                                   &$1.04 \pm 0.04 \pm 0.07$  \\
(b)    &$q^{2}$       &  $E^{B}_{\ell}>1$ GeV,  $M_{X}<1.7$ GeV, $q^{2}>8$ GeV$^{2}$            & $0.70 \pm 0.06 \pm 0.09$  \\
(c1)  & $E^{B}_{\ell}$   &  $E^{B}_{\ell}>1$ GeV, $M_{X}<1.7$ GeV                   &$1.09 \pm 0.05 \pm 0.10$  \\ 
(c2)  & $E^{B}_{\ell}$   &  $E^{B}_{\ell}>1$ GeV                                                         &$1.66 \pm 0.08 \pm 0.17$       \\
(d) & $M_{X}-q^{2}$  &  $E^{B}_{\ell}>1$ GeV                                                                           &$1.56 \pm 0.06 \pm 0.12$    \\ \hline\hline
\end{tabular}
\caption{The measured partial branching fractions for various phase-space regions. The first uncertainty is statistical and the second one is systematics.}
    \label{tab:fit-PS-data}
\end{table}

A binned likelihood fit is performed to extract the signal yield, where the systematic uncertainties are incorporated via nuisance-parameter constraints. The fit uses MC templates for background, and for signal in and out-side of the selected phase-space regions. In total, we carry out five separate fits to measure the three partial branching fractions as summarised in Table~\ref{tab:fit-PS-data}. Fig.~\ref{fig:post-fit} shows the main fit results. The result based on the two-dimensional fit of $M_{X}$ and $q^{2}$, i.e. $\Delta \texttt{BF} = (1.56 \pm 0.06 \pm 0.12) \times 10^{-3}$, is in a good agreement with the one obtained by fitting the lepton spectrum, covering the same phase-space region. It also agrees well with the most precise measurement to date of this region ~\cite{TheBABAR:2016lja}, where $\Delta \texttt{BF} = (1.55 \pm 0.12) \times 10^{-3}$. For other phase-space regions, the measured partial branching fractions are also compatible with the previous measurements \cite{Lees:2011fv}.

Based on the measured partial branching fractions, we calculate the $|V_{ub}|$ value with the theoretical input of decay rate as
\begin{equation}
\left|V_{u b}\right|=\sqrt{\frac{\Delta \mathcal{B}\left(B \rightarrow X_{u} \ell \nu\right)}{\tau_{B} \; \Delta \Gamma\left(B \rightarrow X_{u} \ell \nu\right)}} \;,
\end{equation}
where the average $B$ meson lifetime is taken as $(1.579\pm0.004)$ ps \cite{pdg:2020} and the state-of-the-art theory predictions on $\Delta \Gamma$ are listed in Table~\ref{tab:theory-rates}. Table~\ref{tab:vub} summarises the measured $|V_{ub}|$ values. To quote a single value for $|V_{ub}|$ we adapt the procedure of Ref.~\cite{pdg:2020} and calculate a simple arithmetic average of the most precise determinations for the phase-space region $E^{B}_{\ell}>1$ GeV, obtaining 
\begin{equation}
\left|V_{u b}\right|=(4.06 \pm 0.09 \pm 0.16 \pm 0.15) \times 10^{-3}. 
\end{equation}
\noindent This value is smaller than the previous inclusive measurements of $|V_{ub}|$ in Ref.~\cite{Urquijo:2009tp, Lees:2011fv}. The compatibility with the exclusive measurement of $|V_{ub}|$ in Eq.\ref{eq:avr-vub} is 1.4 standard deviations; it is also compatible with the value expected from CKM unitarity from a global fit of Ref.~\cite{Charles:2004jd} of $\left|V_{u b}\right|=(3.62^{+0.11}_{-0.08}) \times 10^{-3}$ within 1.6 standard deviations.

\begin{table}[t]
\renewcommand\arraystretch{1.2}
\centering
\begin{tabular}{lcccc}
\hline\hline  Phase-space region & BLNP \cite{BLNP} & DGE \cite{DGE1,DGE2}   & GGOU \cite{GGOU}  & ADFR \cite{ADFR1,ADFR2}  \\
\hline$M_{X}<1.7 \mathrm{GeV}$ & $45.2_{-4.6}^{+5.4}$ & $42.3_{-3.8}^{+5.8}$ & $43.7_{-3.2}^{+3.9}$ & $52.3_{-4.7}^{+5.4}$ \\
$M_{X}<1.7 \mathrm{GeV}, q^{2}>8 \mathrm{GeV}^{2}$ & $23.4_{-2.6}^{+3.4}$ & $24.3_{-1.9}^{+2.6}$ & $23.3_{-2.4}^{+3.2}$ & $31.1_{-2.6}^{+3.0}$ \\
$E_{\ell}^{B}>1 \mathrm{GeV}$ & $61.5_{-5.1}^{+6.4}$ & $58.2_{-3.0}^{+3.6}$ & $58.5_{-2.3}^{+2.7}$ & $61.5_{-5.1}^{+5.8}$ \\
\hline \hline
\end{tabular}
\caption{The theory predicted decay rates in the three phase-space regions ($\texttt{ps}^{-1}$).}
    \label{tab:theory-rates}
\end{table}

\begin{table}[t]
\renewcommand\arraystretch{1.5}
\centering
\begin{tabular}{lcccc}
\hline \hline 
Fit   & $10^{3}|V_{ub}|^{\texttt{BLNP}}$ & $10^{3}|V_{ub}|^{\texttt{DGE}}$  & $10^{3}|V_{ub}|^{\texttt{GGOU}}$ & $10^{3}|V_{ub}|^{\texttt{ADFR}}$                      \\
\hline 
(a)  &$3.81_{-0.08,-0.13,-0.21}^{+0.08,+0.13,+0.21}$ & $3.99_{-0.08,-0.14,-0.26}^{+0.08,+0.14,+0.20}$ & $3.88_{-0.08,-0.14,-0.16}^{+0.08,+0.13,+0.15}$ & $3.55_{-0.07,-0.12,-0.17}^{+0.07,+0.12,+0.17}$ \\
(b)  &$4.35_{-0.18,-0.28,-0.28}^{+0.18,+0.26,+0.26}$ & $4.27_{-0.18,-0.28,-0.21}^{+0.17,+0.26,+0.18}$ & $4.36_{-0.18,-0.28,-0.27}^{+0.18,+0.27,+0.24}$ & $3.77_{-0.16,-0.24,-0.17}^{+0.15,+0.23,+0.17}$ \\
(c1) & $3.90_{-0.10,-0.18,-0.21}^{+0.09,+0.17,+0.21}$ & $4.08_{-0.10,-0.19,-0.26}^{+0.10,+0.18,+0.20}$ & $3.97_{-0.10,-0.19,-0.16}^{+0.09,+0.18,+0.15}$ & $3.63_{-0.09,-0.17,-0.17}^{+0.09,+0.16,+0.17}$ \\
(c2) & $4.14_{-0.10,-0.22,-0.20}^{+0.10,+0.20,+0.18}$ & $4.25_{-0.10,-0.22,-0.12}^{+0.10,+0.21,+0.11}$ & $4.24_{-0.10,-0.22,-0.10}^{+0.10,+0.21,+0.09}$ & $4.14_{-0.10,-0.22,-0.18}^{+0.10,+0.20,+0.18}$     \\
(d)  & $4.01_{-0.08,-0.16,-0.19}^{+0.08,+0.15,+0.18}$ & $4.12_{-0.09,-0.16,-0.12}^{+0.08,+0.16,+0.11}$ & $4.11_{-0.09,-0.16,-0.09}^{+0.08,+0.16,+0.08}$ & $4.01_{-0.08,-0.16,-0.18}^{+0.08,+0.15,+0.18}$\\
\hline \hline 
\end{tabular}
\caption{The extracted $|V_{ub}|$ values based on four theoretical inputs on the decay rates. The first uncertainty is statistical, the second one is systematic and the last term comes from the corresponding theory calculation.}
    \label{tab:vub}
\end{table}

\section{Summary and outlook}

The preliminary results are obtained with the hadronic tagged analysis based on the full Belle data set. The measured partial branching fractions for the three phase-space regions are compatible with the previous measurements. The preliminary $|V_{ub}|$ value extracted in this analysis is larger but compatible with the exclusive determination within 1.4 standard deviations. Based on this preliminary result, the final analysis will incorporate a few modifications, including the aspects of increasing the simulated sample size and considering additional systematics accounting for the signal modeling. The separate-mode branching fractions for $B^{+}/B^{0}$ and $e/\mu$ will be also provided.

\bibliographystyle{apsrev4-1}
\bibliography{ref}

\begin{thebibliography}{25}%
\makeatletter
\providecommand \@ifxundefined [1]{%
 \@ifx{#1\undefined}
}%
\providecommand \@ifnum [1]{%
 \ifnum #1\expandafter \@firstoftwo
 \else \expandafter \@secondoftwo
 \fi
}%
\providecommand \@ifx [1]{%
 \ifx #1\expandafter \@firstoftwo
 \else \expandafter \@secondoftwo
 \fi
}%
\providecommand \natexlab [1]{#1}%
\providecommand \enquote  [1]{``#1''}%
\providecommand \bibnamefont  [1]{#1}%
\providecommand \bibfnamefont [1]{#1}%
\providecommand \citenamefont [1]{#1}%
\providecommand \href@noop [0]{\@secondoftwo}%
\providecommand \href [0]{\begingroup \@sanitize@url \@href}%
\providecommand \@href[1]{\@@startlink{#1}\@@href}%
\providecommand \@@href[1]{\endgroup#1\@@endlink}%
\providecommand \@sanitize@url [0]{\catcode `\\12\catcode `\$12\catcode
  `\&12\catcode `\#12\catcode `\^12\catcode `\_12\catcode `\%12\relax}%
\providecommand \@@startlink[1]{}%
\providecommand \@@endlink[0]{}%
\providecommand \url  [0]{\begingroup\@sanitize@url \@url }%
\providecommand \@url [1]{\endgroup\@href {#1}{\urlprefix }}%
\providecommand \urlprefix  [0]{URL }%
\providecommand \Eprint [0]{\href }%
\providecommand \doibase [0]{http://dx.doi.org/}%
\providecommand \selectlanguage [0]{\@gobble}%
\providecommand \bibinfo  [0]{\@secondoftwo}%
\providecommand \bibfield  [0]{\@secondoftwo}%
\providecommand \translation [1]{[#1]}%
\providecommand \BibitemOpen [0]{}%
\providecommand \bibitemStop [0]{}%
\providecommand \bibitemNoStop [0]{.\EOS\space}%
\providecommand \EOS [0]{\spacefactor3000\relax}%
\providecommand \BibitemShut  [1]{\csname bibitem#1\endcsname}%
\let\auto@bib@innerbib\@empty
\bibitem [{\citenamefont {Cabibbo}(1963)}]{Cabibbo:1963yz}%
  \BibitemOpen
  \bibfield  {author} {\bibinfo {author} {\bibfnamefont {N.}~\bibnamefont
  {Cabibbo}},\ }\href {\doibase 10.1103/PhysRevLett.10.531} {\bibfield
  {journal} {\bibinfo  {journal} {Phys. Rev. Lett.}\ }\textbf {\bibinfo
  {volume} {10}},\ \bibinfo {pages} {531} (\bibinfo {year} {1963})}\BibitemShut
  {NoStop}%
\bibitem [{\citenamefont {Kobayashi}\ and\ \citenamefont
  {Maskawa}(1973)}]{Kobayashi:1973fv}%
  \BibitemOpen
  \bibfield  {author} {\bibinfo {author} {\bibfnamefont {M.}~\bibnamefont
  {Kobayashi}}\ and\ \bibinfo {author} {\bibfnamefont {T.}~\bibnamefont
  {Maskawa}},\ }\href {\doibase 10.1143/PTP.49.652} {\bibfield  {journal}
  {\bibinfo  {journal} {Prog. Theor. Phys.}\ }\textbf {\bibinfo {volume}
  {49}},\ \bibinfo {pages} {652} (\bibinfo {year} {1973})}\BibitemShut
  {NoStop}%
\bibitem [{\citenamefont {Amhis}\ \emph {et~al.}(2019)\citenamefont {Amhis}
  \emph {et~al.}}]{Amhis:2019ckw}%
  \BibitemOpen
  \bibfield  {author} {\bibinfo {author} {\bibfnamefont {Y.~S.}\ \bibnamefont
  {Amhis}} \emph {et~al.} (\bibinfo {collaboration} {HFLAV}),\ }\href@noop {}
  {\  (\bibinfo {year} {2019})},\ \Eprint {http://arxiv.org/abs/1909.12524}
  {arXiv:1909.12524 [hep-ex]} \BibitemShut {NoStop}%
\bibitem [{\citenamefont {Abashian}\ \emph {et~al.}(2002)\citenamefont
  {Abashian} \emph {et~al.}}]{Abashian:2000cg}%
  \BibitemOpen
  \bibfield  {author} {\bibinfo {author} {\bibfnamefont {A.}~\bibnamefont
  {Abashian}} \emph {et~al.},\ }\href {\doibase 10.1016/S0168-9002(01)02013-7}
  {\bibfield  {journal} {\bibinfo  {journal} {Nucl. Instrum. Meth.}\ }\textbf
  {\bibinfo {volume} {A479}},\ \bibinfo {pages} {117} (\bibinfo {year}
  {2002})},\ \bibinfo {note} {also see detector section in J.~Brodzicka {\it et
  al.}, Prog. Theor. Exp. Phys. {\bf 2012}, 04D001 (2012).}\BibitemShut {Stop}%
\bibitem [{\citenamefont {Kurokawa}\ and\ \citenamefont
  {Kikutani}(2003)}]{KEKB}%
  \BibitemOpen
  \bibfield  {author} {\bibinfo {author} {\bibfnamefont {S.}~\bibnamefont
  {Kurokawa}}\ and\ \bibinfo {author} {\bibfnamefont {E.}~\bibnamefont
  {Kikutani}},\ }\href {\doibase https://doi.org/10.1016/S0168-9002(02)01771-0}
  {\bibfield  {journal} {\bibinfo  {journal} {Nucl. Instr. and. Meth.}\
  }\textbf {\bibinfo {volume} {A499}},\ \bibinfo {pages} {1 } (\bibinfo {year}
  {2003})},\ \bibinfo {note} {{and other papers included in this Volume; T.~Abe
  {\it et al.}, Prog. Theor. Exp. Phys. {\bf 2013}, 03A001 (2013) and
  references therein.}}\BibitemShut {Stop}%
\bibitem [{\citenamefont {Lange}(2001)}]{EvtGen}%
  \BibitemOpen
  \bibfield  {author} {\bibinfo {author} {\bibfnamefont {D.~J.}\ \bibnamefont
  {Lange}},\ }\href
  {http://www.sciencedirect.com/science/article/pii/S0168900201000894}
  {\bibfield  {journal} {\bibinfo  {journal} {Nucl. Instr. and. Meth.}\
  }\textbf {\bibinfo {volume} {A462}},\ \bibinfo {pages} {152 } (\bibinfo
  {year} {2001})}\BibitemShut {NoStop}%
\bibitem [{\citenamefont {Brun}\ \emph {et~al.}(1987)\citenamefont {Brun},
  \citenamefont {Bruyant}, \citenamefont {Maire}, \citenamefont {McPherson},\
  and\ \citenamefont {Zanarini}}]{Geant3}%
  \BibitemOpen
  \bibfield  {author} {\bibinfo {author} {\bibfnamefont {R.}~\bibnamefont
  {Brun}}, \bibinfo {author} {\bibfnamefont {F.}~\bibnamefont {Bruyant}},
  \bibinfo {author} {\bibfnamefont {M.}~\bibnamefont {Maire}}, \bibinfo
  {author} {\bibfnamefont {A.~C.}\ \bibnamefont {McPherson}}, \ and\ \bibinfo
  {author} {\bibfnamefont {P.}~\bibnamefont {Zanarini}},\ }\href
  {http://inspirehep.net/record/252007?ln=en} {\bibfield  {journal} {\bibinfo
  {journal} {{CERN-DD-EE-84-1}}\ } (\bibinfo {year} {1987})}\BibitemShut
  {NoStop}%
\bibitem [{\citenamefont {Ramirez}\ \emph {et~al.}(1990)\citenamefont
  {Ramirez}, \citenamefont {Donoghue},\ and\ \citenamefont {Burdman}}]{hybrid}%
  \BibitemOpen
  \bibfield  {author} {\bibinfo {author} {\bibfnamefont {C.}~\bibnamefont
  {Ramirez}}, \bibinfo {author} {\bibfnamefont {J.~F.}\ \bibnamefont
  {Donoghue}}, \ and\ \bibinfo {author} {\bibfnamefont {G.}~\bibnamefont
  {Burdman}},\ }\href {\doibase 10.1103/PhysRevD.41.1496} {\bibfield  {journal}
  {\bibinfo  {journal} {Phys. Rev.}\ }\textbf {\bibinfo {volume} {D 41}},\
  \bibinfo {pages} {1496} (\bibinfo {year} {1990})}\BibitemShut {NoStop}%
\bibitem [{\citenamefont {Prim}\ \emph {et~al.}(2020)\citenamefont {Prim} \emph
  {et~al.}}]{Prim:2019gtj}%
  \BibitemOpen
  \bibfield  {author} {\bibinfo {author} {\bibfnamefont {M.}~\bibnamefont
  {Prim}} \emph {et~al.} (\bibinfo {collaboration} {Belle Collaboration}),\
  }\href {\doibase 10.1103/PhysRevD.101.032007} {\bibfield  {journal} {\bibinfo
   {journal} {Phys. Rev. D}\ }\textbf {\bibinfo {volume} {101}},\ \bibinfo
  {pages} {032007} (\bibinfo {year} {2020})}\BibitemShut {NoStop}%
\bibitem [{\citenamefont {De~Fazio}\ and\ \citenamefont
  {Neubert}(1999)}]{DeFazio:1999ptt}%
  \BibitemOpen
  \bibfield  {author} {\bibinfo {author} {\bibfnamefont {F.}~\bibnamefont
  {De~Fazio}}\ and\ \bibinfo {author} {\bibfnamefont {M.}~\bibnamefont
  {Neubert}},\ }\href {\doibase 10.1088/1126-6708/1999/06/017} {\bibfield
  {journal} {\bibinfo  {journal} {JHEP}\ }\textbf {\bibinfo {volume} {06}},\
  \bibinfo {pages} {017} (\bibinfo {year} {1999})}\BibitemShut {NoStop}%
\bibitem [{\citenamefont {Buchmuller}\ and\ \citenamefont
  {Flacher}(2006)}]{Buchmuller:2005zv}%
  \BibitemOpen
  \bibfield  {author} {\bibinfo {author} {\bibfnamefont {O.}~\bibnamefont
  {Buchmuller}}\ and\ \bibinfo {author} {\bibfnamefont {H.}~\bibnamefont
  {Flacher}},\ }\href {\doibase 10.1103/PhysRevD.73.073008} {\bibfield
  {journal} {\bibinfo  {journal} {Phys. Rev. D}\ }\textbf {\bibinfo {volume}
  {73}},\ \bibinfo {pages} {073008} (\bibinfo {year} {2006})},\ \Eprint
  {http://arxiv.org/abs/hep-ph/0507253} {arXiv:hep-ph/0507253} \BibitemShut
  {NoStop}%
\bibitem [{\citenamefont {Feindt}\ \emph {et~al.}(2011)\citenamefont {Feindt},
  \citenamefont {Keller}, \citenamefont {Kreps}, \citenamefont {Kuhr},
  \citenamefont {Neubauer}, \citenamefont {Zander},\ and\ \citenamefont
  {Zupanc}}]{Feindt:2011mr}%
  \BibitemOpen
  \bibfield  {author} {\bibinfo {author} {\bibfnamefont {M.}~\bibnamefont
  {Feindt}}, \bibinfo {author} {\bibfnamefont {F.}~\bibnamefont {Keller}},
  \bibinfo {author} {\bibfnamefont {M.}~\bibnamefont {Kreps}}, \bibinfo
  {author} {\bibfnamefont {T.}~\bibnamefont {Kuhr}}, \bibinfo {author}
  {\bibfnamefont {S.}~\bibnamefont {Neubauer}}, \bibinfo {author}
  {\bibfnamefont {D.}~\bibnamefont {Zander}}, \ and\ \bibinfo {author}
  {\bibfnamefont {A.}~\bibnamefont {Zupanc}},\ }\href {\doibase
  10.1016/j.nima.2011.06.008} {\bibfield  {journal} {\bibinfo  {journal} {Nucl.
  Instrum. Meth. A}\ }\textbf {\bibinfo {volume} {654}},\ \bibinfo {pages}
  {432} (\bibinfo {year} {2011})}\BibitemShut {NoStop}%
\bibitem [{\citenamefont {Bevan}\ \emph {et~al.}(e 95)\citenamefont {Bevan}
  \emph {et~al.}}]{Bevan:2014iga}%
  \BibitemOpen
  \bibfield  {author} {\bibinfo {author} {\bibfnamefont {A.}~\bibnamefont
  {Bevan}} \emph {et~al.},\ }\href {\doibase 10.1140/epjc/s10052-014-3026-9}
  {\bibfield  {journal} {\bibinfo  {journal} {Eur. Phys. J. C}\ }\textbf
  {\bibinfo {volume} {74}},\ \bibinfo {pages} {3026} (\bibinfo {year} {2014,
  Page 95})}\BibitemShut {NoStop}%
\bibitem [{\citenamefont {Glattauer}\ \emph {et~al.}(2016)\citenamefont
  {Glattauer} \emph {et~al.}}]{Glattauer:2015teq}%
  \BibitemOpen
  \bibfield  {author} {\bibinfo {author} {\bibfnamefont {R.}~\bibnamefont
  {Glattauer}} \emph {et~al.} (\bibinfo {collaboration} {Belle
  Collaboration}),\ }\href {\doibase 10.1103/PhysRevD.93.032006} {\bibfield
  {journal} {\bibinfo  {journal} {Phys. Rev.}\ }\textbf {\bibinfo {volume}
  {D93}},\ \bibinfo {pages} {032006} (\bibinfo {year} {2016})}\BibitemShut
  {NoStop}%
\bibitem [{\citenamefont {Lees}\ \emph {et~al.}(2017)\citenamefont {Lees} \emph
  {et~al.}}]{TheBABAR:2016lja}%
  \BibitemOpen
  \bibfield  {author} {\bibinfo {author} {\bibfnamefont {J.}~\bibnamefont
  {Lees}} \emph {et~al.} (\bibinfo {collaboration} {BaBar Collaboration}),\
  }\href {\doibase 10.1103/PhysRevD.95.072001} {\bibfield  {journal} {\bibinfo
  {journal} {Phys. Rev. D}\ }\textbf {\bibinfo {volume} {95}},\ \bibinfo
  {pages} {072001} (\bibinfo {year} {2017})}\BibitemShut {NoStop}%
\bibitem [{\citenamefont {Lees}\ \emph {et~al.}(2012)\citenamefont {Lees} \emph
  {et~al.}}]{Lees:2011fv}%
  \BibitemOpen
  \bibfield  {author} {\bibinfo {author} {\bibfnamefont {J.}~\bibnamefont
  {Lees}} \emph {et~al.} (\bibinfo {collaboration} {BaBar Collaboration}),\
  }\href {\doibase 10.1103/PhysRevD.86.032004} {\bibfield  {journal} {\bibinfo
  {journal} {Phys. Rev. D}\ }\textbf {\bibinfo {volume} {86}},\ \bibinfo
  {pages} {032004} (\bibinfo {year} {2012})}\BibitemShut {NoStop}%
\bibitem [{\citenamefont {Zyla}\ \emph {et~al.}(2020)\citenamefont {Zyla} \emph
  {et~al.}}]{pdg:2020}%
  \BibitemOpen
  \bibfield  {author} {\bibinfo {author} {\bibfnamefont {P.}~\bibnamefont
  {Zyla}} \emph {et~al.} (\bibinfo {collaboration} {Particle Data Group}),\
  }\href@noop {} {\bibfield  {journal} {\bibinfo  {journal} {Prog. Theor. Exp.
  Phys. {\bf 2020}}\ }\textbf {\bibinfo {volume} {083C01}} (\bibinfo {year}
  {2020})}\BibitemShut {NoStop}%
\bibitem [{\citenamefont {Urquijo}\ \emph {et~al.}(2010)\citenamefont {Urquijo}
  \emph {et~al.}}]{Urquijo:2009tp}%
  \BibitemOpen
  \bibfield  {author} {\bibinfo {author} {\bibfnamefont {P.}~\bibnamefont
  {Urquijo}} \emph {et~al.} (\bibinfo {collaboration} {Belle Collaboration}),\
  }\href {\doibase 10.1103/PhysRevLett.104.021801} {\bibfield  {journal}
  {\bibinfo  {journal} {Phys. Rev. Lett.}\ }\textbf {\bibinfo {volume} {104}},\
  \bibinfo {pages} {021801} (\bibinfo {year} {2010})}\BibitemShut {NoStop}%
\bibitem [{\citenamefont {Charles}\ \emph {et~al.}(2005)\citenamefont
  {Charles}, \citenamefont {Hocker}, \citenamefont {Lacker}, \citenamefont
  {Laplace}, \citenamefont {Le~Diberder}, \citenamefont {Malcles},
  \citenamefont {Ocariz}, \citenamefont {Pivk},\ and\ \citenamefont
  {Roos}}]{Charles:2004jd}%
  \BibitemOpen
  \bibfield  {author} {\bibinfo {author} {\bibfnamefont {J.}~\bibnamefont
  {Charles}}, \bibinfo {author} {\bibfnamefont {A.}~\bibnamefont {Hocker}},
  \bibinfo {author} {\bibfnamefont {H.}~\bibnamefont {Lacker}}, \bibinfo
  {author} {\bibfnamefont {S.}~\bibnamefont {Laplace}}, \bibinfo {author}
  {\bibfnamefont {F.}~\bibnamefont {Le~Diberder}}, \bibinfo {author}
  {\bibfnamefont {J.}~\bibnamefont {Malcles}}, \bibinfo {author} {\bibfnamefont
  {J.}~\bibnamefont {Ocariz}}, \bibinfo {author} {\bibfnamefont
  {M.}~\bibnamefont {Pivk}}, \ and\ \bibinfo {author} {\bibfnamefont
  {L.}~\bibnamefont {Roos}} (\bibinfo {collaboration} {CKMfitter Group}),\
  }\href {\doibase 10.1140/epjc/s2005-02169-1} {\bibfield  {journal} {\bibinfo
  {journal} {Eur. Phys. J. C}\ }\textbf {\bibinfo {volume} {41}},\ \bibinfo
  {pages} {1} (\bibinfo {year} {2005})}\BibitemShut {NoStop}%
\bibitem [{\citenamefont {Lange}\ \emph {et~al.}(2005)\citenamefont {Lange},
  \citenamefont {Neubert},\ and\ \citenamefont {Paz}}]{BLNP}%
  \BibitemOpen
  \bibfield  {author} {\bibinfo {author} {\bibfnamefont {B.~O.}\ \bibnamefont
  {Lange}}, \bibinfo {author} {\bibfnamefont {M.}~\bibnamefont {Neubert}}, \
  and\ \bibinfo {author} {\bibfnamefont {G.}~\bibnamefont {Paz}},\ }\href
  {\doibase 10.1103/PhysRevD.72.073006} {\bibfield  {journal} {\bibinfo
  {journal} {Phys. Rev. D}\ }\textbf {\bibinfo {volume} {72}},\ \bibinfo
  {pages} {073006} (\bibinfo {year} {2005})}\BibitemShut {NoStop}%
\bibitem [{\citenamefont {Andersen}\ and\ \citenamefont {Gardi}(2006)}]{DGE1}%
  \BibitemOpen
  \bibfield  {author} {\bibinfo {author} {\bibfnamefont {J.~R.}\ \bibnamefont
  {Andersen}}\ and\ \bibinfo {author} {\bibfnamefont {E.}~\bibnamefont
  {Gardi}},\ }\href {\doibase 10.1088/1126-6708/2006/01/097} {\bibfield
  {journal} {\bibinfo  {journal} {JHEP}\ }\textbf {\bibinfo {volume} {01}},\
  \bibinfo {pages} {097} (\bibinfo {year} {2006})}\BibitemShut {NoStop}%
\bibitem [{\citenamefont {Gardi}(2008)}]{DGE2}%
  \BibitemOpen
  \bibfield  {author} {\bibinfo {author} {\bibfnamefont {E.}~\bibnamefont
  {Gardi}},\ }\href@noop {} {\bibfield  {journal} {\bibinfo  {journal}
  {Frascati Phys. Ser.}\ }\textbf {\bibinfo {volume} {47}},\ \bibinfo {pages}
  {381} (\bibinfo {year} {2008})}\BibitemShut {NoStop}%
\bibitem [{\citenamefont {Gambino}\ \emph {et~al.}(2007)\citenamefont
  {Gambino}, \citenamefont {Giordano}, \citenamefont {Ossola},\ and\
  \citenamefont {Uraltsev}}]{GGOU}%
  \BibitemOpen
  \bibfield  {author} {\bibinfo {author} {\bibfnamefont {P.}~\bibnamefont
  {Gambino}}, \bibinfo {author} {\bibfnamefont {P.}~\bibnamefont {Giordano}},
  \bibinfo {author} {\bibfnamefont {G.}~\bibnamefont {Ossola}}, \ and\ \bibinfo
  {author} {\bibfnamefont {N.}~\bibnamefont {Uraltsev}},\ }\href {\doibase
  10.1088/1126-6708/2007/10/058} {\bibfield  {journal} {\bibinfo  {journal}
  {JHEP}\ }\textbf {\bibinfo {volume} {10}},\ \bibinfo {pages} {058} (\bibinfo
  {year} {2007})}\BibitemShut {NoStop}%
\bibitem [{\citenamefont {Aglietti}\ \emph {et~al.}(2009)\citenamefont
  {Aglietti}, \citenamefont {Di~Lodovico}, \citenamefont {Ferrera},\ and\
  \citenamefont {Ricciardi}}]{ADFR1}%
  \BibitemOpen
  \bibfield  {author} {\bibinfo {author} {\bibfnamefont {U.}~\bibnamefont
  {Aglietti}}, \bibinfo {author} {\bibfnamefont {F.}~\bibnamefont
  {Di~Lodovico}}, \bibinfo {author} {\bibfnamefont {G.}~\bibnamefont
  {Ferrera}}, \ and\ \bibinfo {author} {\bibfnamefont {G.}~\bibnamefont
  {Ricciardi}},\ }\href {\doibase 10.1140/epjc/s10052-008-0817-x} {\bibfield
  {journal} {\bibinfo  {journal} {Eur. Phys. J. C}\ }\textbf {\bibinfo {volume}
  {59}},\ \bibinfo {pages} {831} (\bibinfo {year} {2009})}\BibitemShut
  {NoStop}%
\bibitem [{\citenamefont {Aglietti}\ \emph {et~al.}(2007)\citenamefont
  {Aglietti}, \citenamefont {Ferrera},\ and\ \citenamefont
  {Ricciardi}}]{ADFR2}%
  \BibitemOpen
  \bibfield  {author} {\bibinfo {author} {\bibfnamefont {U.}~\bibnamefont
  {Aglietti}}, \bibinfo {author} {\bibfnamefont {G.}~\bibnamefont {Ferrera}}, \
  and\ \bibinfo {author} {\bibfnamefont {G.}~\bibnamefont {Ricciardi}},\ }\href
  {\doibase 10.1016/j.nuclphysb.2007.01.014} {\bibfield  {journal} {\bibinfo
  {journal} {Nucl. Phys. B}\ }\textbf {\bibinfo {volume} {768}},\ \bibinfo
  {pages} {85} (\bibinfo {year} {2007})}\BibitemShut {NoStop}%
\end{thebibliography}%

\end{document}